\documentclass[twocolumn,usenatbib,prl,floats,usegraphicx]{revtex4}

\usepackage{epsfig}
\usepackage{graphicx}
\usepackage{graphics}

\usepackage{amsfonts}  
\usepackage{amsmath, amssymb, stmaryrd}     

\def\be{\begin{equation}}
\def\ee{\end{equation}}
\def\bi{\begin{itemize}}
\def\ei{\end{itemize}}

\newcommand\mnras{\ref@jnl{MNRAS}}%
\newcommand\aap{\ref@jnl{A\&A}}%
\begin{document}

\title{Releasing scalar fields:  cosmological simulations of scalar-tensor theories for gravity beyond the static approximation}

\author{Claudio Llinares}
\affiliation{Institute of Theoretical Astrophysics, University of Oslo, N-0315 Oslo, Norway}
\author{David F. Mota}
\affiliation{Institute of Theoretical Astrophysics, University of Oslo, N-0315 Oslo, Norway}

\begin{abstract}
Several extensions of General Relativity and high energy physics include scalar fields as extra degrees of freedom.  In the search for predictions in the non-linear regime of cosmological evolution, the community makes use of numerical simulations in which the quasi-static limit is assumed when solving the equation of motion of the scalar field.  In this Letter, we propose a method to solve the full equations of motion for scalar degrees of freedom coupled to matter. We run cosmological simulations which track the full time and space evolution of the scalar field, and find striking differences with respect to the commonly used quasi-static approximation. This novel procedure reveals new physical properties of the scalar field and uncovers concealed astrophysical phenomena which were hidden in the old approach. 
\end{abstract}

\maketitle

General Relativity (GR) is the foundation stone of the standard cosmological model.  The assumption that GR describes gravity at cosmological scales leads to two extra building blocks of the model: dark matter and dark energy.  In spite of the model's ability to match a large number of observations, the nature of the two dark components is still unclear. 
Several extensions of General Relativity and the Standard Model of Particle Physics have been proposed to explain the dark sector. Most of this extensions include scalar fields as extra degrees of freedom which interact with matter and lead to modifications to standard gravity \citep[][]{2012PhR...513....1C}. 
As GR is proven to be valid in solar system scales, any modification introduced must fulfil the requirement of reducing to Einstein gravity at small scales. This is assured through screening mechanisms \citep{ 2010PhRvL.104w1301H, cham, 2012PhRvL.109x1102K}. Such processes are highly non-linear, therefore are better probed in the non-linear regime of cosmological structure formation in the scales of galaxies and clusters of galaxies.

With the intention to probe the nature of the dark sector, modifications to GR and the indispensable screening mechanisms, the community makes use of N-body simulations. The main assumption behind such simulations is the quasi-static limit when solving the equation of motion of the extra scalar degrees of freedom  (i.e. neglect time derivatives) \citep[][]{2011arXiv1108.3081D,bao1,baldi,fab1}. The validity of this quasi-static hypothesis was never tested. 

In this Letter, we propose a method to solve the full equations of motion for scalar degrees of freedom coupled to matter. We run cosmological simulations which track the full time and space evolution of the scalar field, and find striking differences with respect to the commonly used quasi-static approximation. Our novel procedure reveals new physical properties of the scalar field and reveals concealed astrophysical phenomena which were hidden in the old approach. We show how such results may have an impact on both the theoretical studies of Modified Gravity and its astrophysical probes. 

We focus on  a general class  of scalar-tensor theories which is characterized by the symmetron screening mechanism \citep[][]{2010PhRvL.104w1301H}.  However, our results and techniques can be straightforwardly generalised to any theory with a scalar degree of freedom, independently of the screening mechanism that is invoked.  Moreover, the numerical methods used are also valid for integrating the equations either in the Einstein or Jordan frame \citep[][]{2011PhR...509..167C}. In our case, solar system constraints imply that the conformal factor relating the two frames is close to unity. Therefore, the solutions in both frames are similar.

The symmetron model is defined by the following effective potential:
\be
V_{s,eff}(\phi) = \frac{1}{2} \left(\frac{\rho}{M^2} - \mu^2 \right) \phi^2 + \frac{1}{4}\lambda \phi^4, 
\ee
which leads to the following equation of motion:
\be
\ddot{\phi} + 3 H \dot{\phi} - c^2\frac{\nabla^2\phi}{a^2} = -c^2\left[\left( \frac{\rho}{M^2} -\mu^2\right)\phi + \lambda\phi^3\right]
\label{eq_motion_phi}
\ee
where $\mu$ and $M$ are mass scales and $\rho$ is the matter density.  We normalize the field $\phi$ with its vacuum expectation value $\phi_{0}^2 = \frac{\mu^2}{\lambda}$.
Defining
$\chi=\frac{\phi}{\phi_{0}}$,
and taking into account that the background density acquires a value  
$\rho_{SSB} = M^2 \mu^2$ at  $z_{SSB}$,
we obtain:
\be
\label{eq_motion_chi}
\ddot{\chi} + 3H\dot{\chi} - c^2\frac{\nabla^2\chi}{a^2} = -\frac{c^2}{2\lambda_0^2} \left[ \frac{a_{SSB}^3}{a^3}\chi\eta - \chi + \chi^3  \right], 
\ee
where $\eta$ is the matter density field normalised with the background density, and $\lambda_{0} = \frac{1}{\sqrt{2} \mu}$ is the range of the field that corresponds to $\eta=0$.

The Newtonian potential $\phi_N$ is given by:
\be
\nabla^2 \phi_N = \frac{3}{2}\frac{\Omega_m H_0^2}{a} \delta, 
\label{poisson_eq}
\ee
where $\delta$ is the over-density defined as $\delta\rho/\rho_b$, $\rho_b$ is the background density, $\phi_N$ is the perturbation in the Einstein frame metric.  The particles that are used to track the dark matter component do not follow geodesics of this metric, but a modified version of it that includes a fifth force associated to the scalar field:
\be
\ddot{\textbf{x}} + 2 H \dot{\textbf{x}} + \frac{1}{a^2}\nabla\left(\phi_N + 6H_0^2 \Omega_m \lambda_{0}^2 \frac{(1+z_{SSB})^3}{2}\beta^2 \chi^2 \right) = 0. 
\label{geodesics}
\ee

The standard way of solving equation (\ref{eq_motion_chi}) is to assume the quasi-static limit and use a standard multigrid method to solve the resulting elliptic equation.  In this Letter we go beyond this approach and solve the complete equation.  Our new method is based on the fact that the equation of motion for the scalar field, eq.(\ref{eq_motion_chi}), is formally equivalent to the geodesics equation, eq.(\ref{geodesics}).  We can, therefore, apply a leap-frog scheme, not to positions and velocities of particles as in an N-body code, but to the scalar field $\chi$ and its time derivative on a grid.  The change 
$q = a^3 \dot{\chi}$ gives the following first order equations:
\begin{align}
& \dot{\chi} = \frac{q}{a^3}\\
\label{hamilton_2}
& \dot{q} = c^2 a \nabla^2\chi - \frac{a^3 c^2}{2\lambda_0^2} \left[ \frac{a_{SSB}^3}{a^3}\chi\eta - \chi + \chi^3  \right].
\end{align}
By using second order discretization in time and implementing a leap frog scheme, we obtain the following equations for the field and the particles for the time step $n$:
\begin{align}
q_{n+1/2} & = q_{n-1/2} + \\
\nonumber
& \left\{\nabla^2\chi_n - \frac{a_n^2}{2\lambda_0^2} \left[ \frac{a_{SSB}^3}{a_n^3}\chi_n\eta_n - \chi_n + \chi_n^3  \right]  \right\}\frac{a_n c^2}{\dot{a}_n}\Delta a \\
v_{n+1/2} & = v_{n-1/2} - \frac{(\nabla\Phi)_n}{\dot{a}_n} \Delta a\\
\chi_{n+1} & = \chi_n + \frac{q_{n+1/2}}{\dot{a}_{n+1/2} a^3_{n+1/2}} \Delta a \\
x_{n+1} & = x_{n} + \frac{v_{n+1/2}}{\dot{a}_{n+1/2}a^2_{n+1/2}} \Delta a, 
\end{align}
where $\Phi$ is the total potential that includes GR and the scalar field and $a$ is employed as time variable.  
This method was successfully applied to compute solutions of the growth equation for linear density perturbations coupled to a non-linear Poisson equation \citep[][]{llinares_thesis}.  Furthermore, the coupling with the geodesics was employed to generate initial conditions for non-linear gravities \citep[][]{llinares_thesis}.  A similar approach was also applied in the context of scalar fields that are not coupled to matter \citep[][]{1989PhRvD..40.1002W}.

The implementation of the solution requires initial conditions for the variables $\chi$ and $q$.  We choose to fix $\chi=0$ at the initial time, which corresponds to a fully screened scalar field as is expected at early times.  For $q$, we use a small random number. 

Our new method and its parallelisation were implemented in the N-body code presented in \citep{llinares_thesis}, to which the leap-frog associated to the time evolution of the scalar field was added.  
The algorithm and the 3D solver were tested satisfactorily against 1D analytic linear solutions that were obtained for a fixed density distribution given by a gaussian profile located in the center of the box.
\begin{figure}
  \begin{center}
    \includegraphics[width=0.45\textwidth]{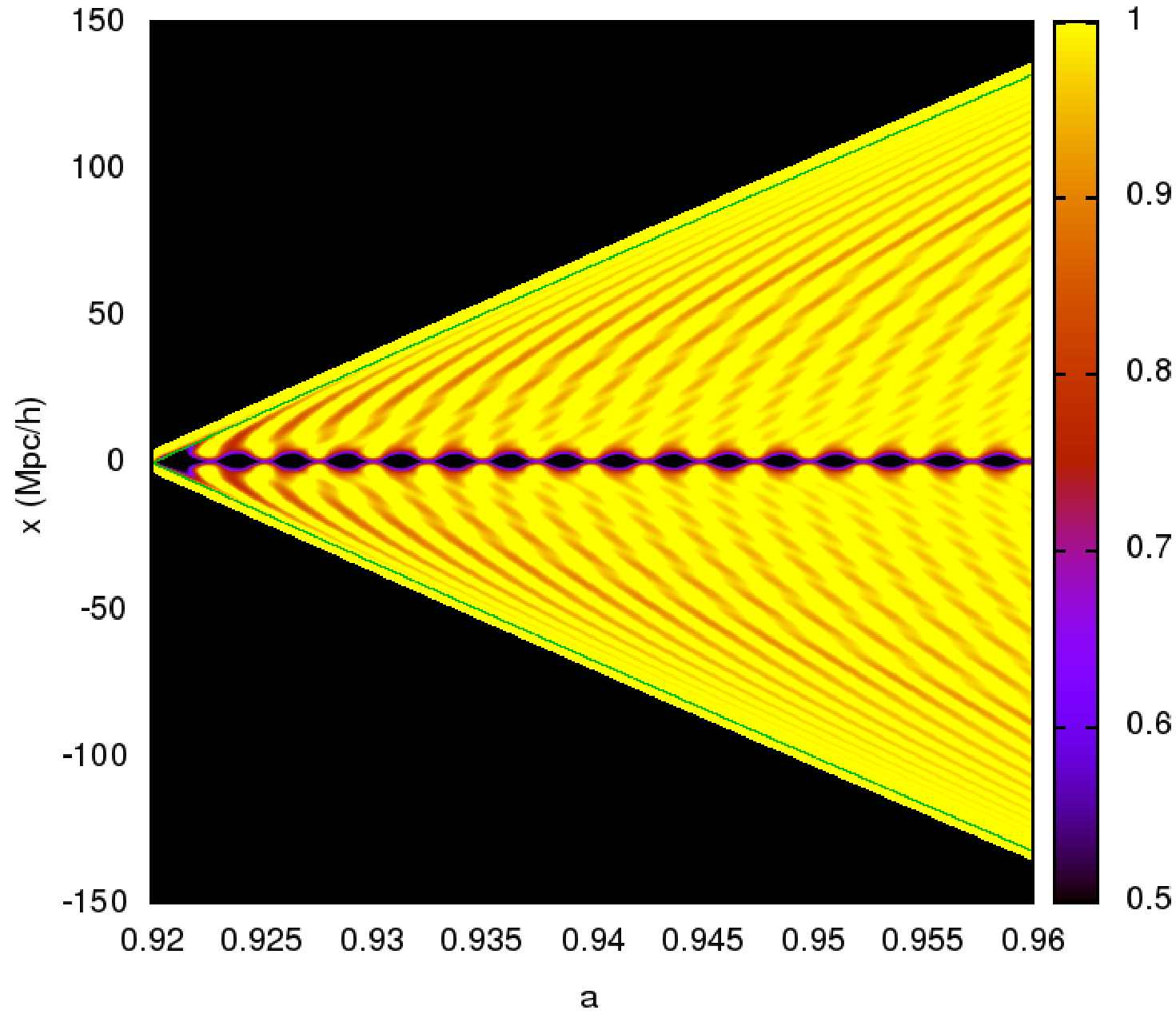}
    \caption{Normalised scalar field $\chi$ as a function of space and time for a fixed Gaussian density distribution.  For clarity, the unperturbed regions where $\chi=1$ are shown in black.  The green lines are null geodesics that pass through the center of the halo at the initial time.  Here, $z_{SSB}=1$ and $\lambda_0=1$ Mpc/h.}
    \label{fig:1d_run}
  \end{center}
\end{figure}
\begin{figure*}
  \begin{minipage}[t]{0.66\textwidth}
    \includegraphics[width=1.0\textwidth]{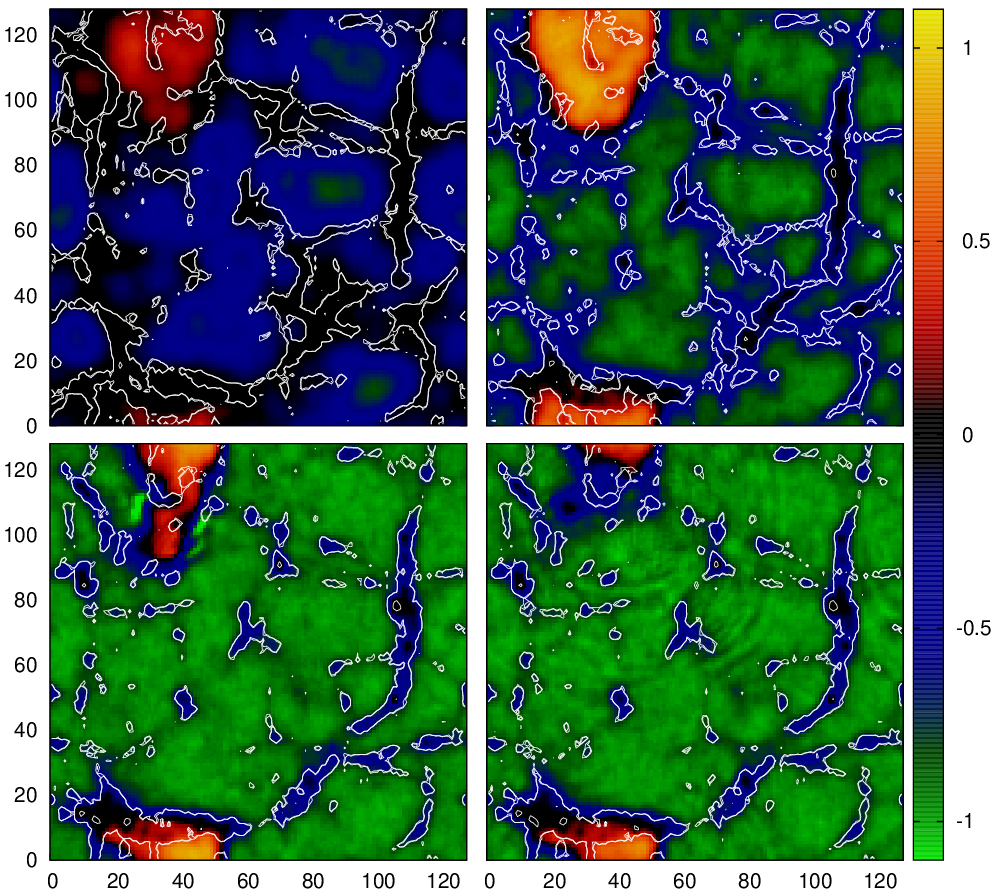}
  \end{minipage}
  \hfill{}
  \begin{minipage}[t]{0.32\textwidth}
    \hspace{-0.7cm}
    \includegraphics[width=1.1\textwidth]{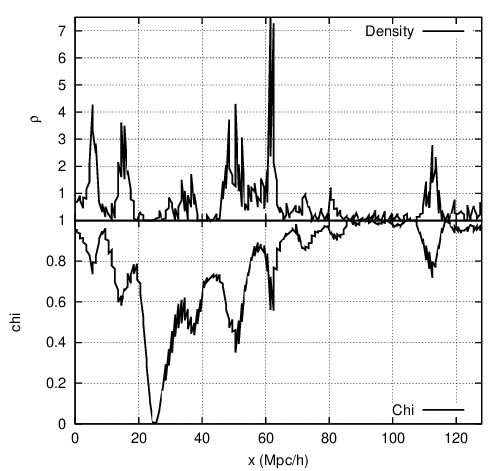}
  \end{minipage}
  \caption{The four color panels show the normalised scalar field $\chi$ for a particular slice of the simulation at four different redshifts ($z=0.807, 0.366, 0.03$ and $0.008$).  The white lines are density contours.  Up-left: scalar field developing in voids ($z_{SSB}=0.5$).  Up-Right: a domain wall which is stable from its formation up to $z\sim 0.03$.  Bottom-left: collapse of the wall.  Bottom-right: a spherical wave originated in that collapse and a dark blob (see the upper-left region where the wall was).  The extra panel to the right shows a cut of the last color panel through a row with $y=108$ Mpc/h (the curves are density and normalised scalar field).  Note the dark blob located in $x\sim 25$ Mpc/h.}
  \label{fig:panels_chi}
\end{figure*}

For the first time in cosmological simulations of modified gravity one can compute the speed at which the new scalar degree of freedom travels. 
To this end, we use the most straightforward definition, which is the velocity at which perturbations propagate in space away from a given source.  The origin of the perturbations is given by a fixed 1D Gaussian density profile with a dispersion of 1 Mpc/h located in the center of the domain.  
The initial conditions are $\chi=1$ and $q=0$, which means that we start the simulations with an unperturbed vacuum field and measure the speed of the perturbations induced by the density as they move away from the source.  A graphical representation of the speed of sound can be seen in Fig.\ref{fig:1d_run}, where we show $\chi$ as a function of space and time.  The bulk of the mass that induce the oscillations is located in the center of the domain.  The green lines are null geodesics that depart from the center of the halo at the initial time.  The figure shows that the front wave of the perturbations in the scalar field moves with the same speed as light, as expected due to its standard kinetic term.  We performed several simulations with different configurations for the underlying density distributions and found no deviation from this value for ranges $\lambda_0$ from $10^{-4}$ to 1 Mpc/h. 

In order to highlight the consequences of taking into account non-static terms in the cosmological evolution of the scalar field, we present results obtained from a cosmological simulation that was run with a box at redshift $z=0$ of 128 Mpc/h and using $128^3$ particles.  The scalar field and gravitational potential were tracked in a uniform grid with 128 nodes per dimension.  As the evolution of the matter density and the scalar field have different time scales and the aim of this Letter is to understand only the evolution of the scalar field, we evolve the matter component by taking into account the GR equations.  The small differences that will appear in the matter component because of considering also the fifth force in its evolution are not expected to change the main properties of the non-static solution that we intend to study here.  The background evolution is given by a flat $\Lambda$CDM cosmology ($\Omega_m=0.267$, $\Omega_{\Lambda}=0.733$ and $H_0=71.9$ km/sec/Mpc).  The initial conditions for matter were generated using Zeldovich approximation with standard gravity with the package Cosmics \citep[][]{1995astro.ph..6070B}.  The particular symmetron parameters employed are $z_{SSB}=0.5$ and $\lambda_0=1$ Mpc/h, which provide a cosmological evolution close to $\Lambda$CDM \citep{2011arXiv1108.3081D}.  

\begin{figure}[!t]
  \begin{center}
    \includegraphics[width=0.45\textwidth]{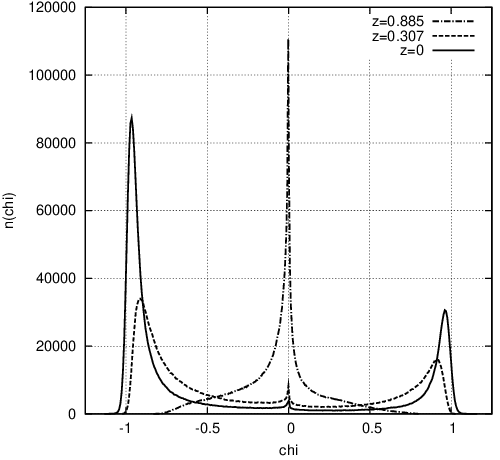}
    \caption{Histograms of normalised scalar field $\chi$ on the grid for three different redshifts of the cosmological simulation.}
    \label{fig:histogram}
  \end{center}
\end{figure}

The time steeping is uniform in $a$ space during the simulations.  We used $10^4$ time steps to evolve GR and the position of the particles.  The evolution of the scalar field has a much shorter time scale than GR and thus, it must be tracked with a much shorter time step.  We made runs with $9.6\times 10^5$ and $1.92\times 10^6$ time steps and found the same statistical distribution of the scalar field (shown in Fig.\ref{fig:histogram}), which shows that this number of steps is enough for the spatial resolution employed.

Fig.\ref{fig:panels_chi} shows the distribution of the non-static scalar field for a given slice of the box at four different redshifts ($z=0.807, 0.366, 0.03$ and $0.008$).  The white lines are contours of the matter density.  The different snapshots were chosen to highlight different properties of the non-static solutions.

The symmetron has an effective potential which acquires two minima when the symmetry is broken and thus, the scalar field can adopt different signs in different regions in space.  In the static cosmological simulations one needs to fix the sign of the field to be unique all over the box (relaxing this constraint make iterative solvers to fail to converge).  As in our simulations the evolution of the scalar field is followed in a self consistent way, we can study the validity of this constraint.  We find that the sign of $\chi$ is indeed not necessarily unique.  On the contrary, after $z_{SSB}$, the box appears divided in different domains in which the scalar field chooses different signs.  We find that the domain walls associated to this decomposition can be stable during long period of time until they collapse.  The upper panels of the Fig.\ref{fig:panels_chi} show the formation of a domain wall in the upper-left part of the box.  This particular wall does not survive until redshift $z=0$.  In the bottom-left panel, it is possible to see the wall collapsing.  The energy that was contained in it is released in the form of scalar waves that can travel several megaparsecs.  The front wave associated to this particular event can be found in the lower-right panel crossing the center of the box.  It is important to note that in the static case, the matter distribution only fixes the boundary conditions.  However, the exact time evolution of $\chi$ is a function not only of the boundary conditions, but also of its history, which is not described by the static equation.  Even if the domain walls can be obtained from the static solution, the static equation does not provide enough information to describe their evolution in time, neither the release of energy and propagation of scalar waves. All these factors affect the behaviour (range and strength) of the fifth force induced by the scalar field and so the formation of cosmological structures. 

The distribution of signs for the simulation presented in this Letter can be seen in Fig.\ref{fig:histogram}, where we show histograms of the scalar field made over the whole grid at different redshifts.  At redshift close to symmetry breaking, the scalar field is almost fully screened and thus, it has a large peak in zero.  Later on, two large picks develop in the distribution, which corresponds to domains with positive and negative $\chi$.  For the particular simulation shown here, the bimodal distribution subsist until $z=0$.  We find that this is not a general result, but that by changing the initial seed of the initial conditions it is possible to obtain distributions completely skewed towards positive or negative values.

The non-linearity of modified gravity can induce a decoupling between dynamical and real mass, i.e. the modified force can converge towards regions devoid of matter \citep[][]{2009ApJ...703.2285K}.  Here we study the possibility that non-static terms can also induce similar effects and find that the answer is indeed positive.  We invite the reader to focus his/her attention in the lower-right color panel of Fig.\ref{fig:panels_chi}.  The figure shows that each feature in the scalar field distribution has a counterpart in the density distribution with exception of the minimum of $|\chi|$ at $(x,y)\sim (25,108)$ Mpc/h (it can be seen as a black region in the color figure).  We find that this feature of the solution of the field equation is not a consequence of an external field as in \citep[][]{2009ApJ...703.2285K}, but a consequence of the non-static terms: the effect does not appear in the static solution that correspond to the same density distribution.  Furthermore, the depression in $\chi$ moves in space, a characteristic that cannot be explained by using the static approximation.  We find the same effect in several simulations, usually associated to domain walls.  While the energy associated with the scalar field is not enough to explain the dark blobs that are present in several clusters of galaxies \citep[][]{{2012ApJ...747...96J},{2007ApJ...668..806M},{2007PhRvD..76d3524R}}, this peculiar characteristic of the non-static solution must be taken into account when developing new observables to detect scalar fields and probe modifications to GR.

In summary, we present a new method to solve the full equation of motion for scalar-tensor theories that does not rely on the quasi-static approximation.  The method is based on a leap-frog scheme and can be generalised in a trivial way to a number of commonly studied modified gravity models by changing the source function in eq.(\ref{hamilton_2}).  In order to test the consequence of the non-static terms on the evolution of the scalar field, we run cosmological simulations with our new solver.  We find several new striking physical features which are not present in the usual static approach. For instance, the scalar degree of freedom develops domain walls whose evolution in time cannot be described by the static equation. The collapse of these domain walls are source of scalar waves that travel across several Mpc and may leave a cosmological imprint in galaxies and clusters at those scales. Finally, our algorithm shows that it is possible to decouple the potential wells in the scalar field from overdense regions, leading to the decoupling of real and dynamical mass.   

Simulations show that matter can fall in the domain walls producing a local increase in the matter density.  This makes the domain walls to leave an imprint in the cosmic shear that should be possible to observe with large lensing surveys such as the incoming ESA Euclid survey.  Furthermore, as these modifications in the density depend on the position of the walls and not on the matter distribution, the non-static terms provide a new source of non-gaussianities that should be taken into account.  While the predictions related to domain walls are specific of the symmetron model, there are other observables related to the oscillations in the scalar field that are model independent.  In this sense, studies of mass accretion in galaxies for instance through $L_{\alpha}$ emission (which occurs in regions where the scalar fields oscillate) and generation of X-rays in shock waves induced by the oscillations will be crucial when searching for tighter constraints for scalar-tensor theories.  The implications of these effects if observed in near future astrophysical probes at galactic and cluster scales would open a new window to test novel gravity theories and high energy models with cosmological simulations of the nonlinear regime.


\bibliography{references}

\end{document}